\documentclass[review]{elsarticle}

\usepackage{lineno,hyperref}
\modulolinenumbers[5]
\usepackage{lscape}
\journal{Journal of \LaTeX\ Templates}

    \def\bea{\begin{eqnarray}}
  \def\eea{\end{eqnarray}}

  \def\be{\begin{dmath}}  \def\ee{\end{dmath}}
\begin{document}

\begin{frontmatter}

\title{Spectroscopy of Heavy-Light Mesons ($c\bar{s}$, $c\bar{q}$, $b\bar{s}$, $b\bar{q}$) for the linear plus modified  Yukawa potential using Nikiforov-Uvarov
Method\tnoteref{mytitlenote}}

\author{ Kaushal R Purohit\fnref{myfootnote}$^*$}
\ead{kaushalsep1996@gmail.com}
\author{Ajay Kumar Rai $^1$}
\author{Rajendrasinh H Parmar$^2$}
\address{Department of Physics, Sardar Vallabhbhai National Institute of Technology, Surat, Gujarat-395 007, India$^1$}
\address{Sir P T Science College, Modasa-383 315, Gujarat India$^2$}



\begin{abstract}
An approximate bound state solution of the Klein-Gordon equation is derive analytically for the 3-dimensional space with a combination framework of linear plus modified Yukawa Potential (LIMYP) using the Nikiforov-Uvarov (N-U) method for obtaining the energy eigenvalues and corresponding wave function. A detailed study of mass spectra of all combination sets of heavy-light flavor mesons vis-a-vis $(Ks/Kq; K= C, B)$ is investigated by treating both heavy-light flavor mesons non-relativistic with an effective quark-antiquark interaction potential for different quantum states. Along with that, an elucidated graphical representation is scrutinized with the calculated mass spectra obtained from the energy eigenvalue against the corresponding variables for all the combination sets of heavy-light flavors mesons. Therefore, the current framework potential provides excellent reconciliation with the experimental data of states known to date and minuscule \%  difference in lower quantum states, which increases with higher quantum states that can be correlated with the higher screening factor coming into the account.
\end{abstract}


\end{frontmatter}

\section{Introduction}

One of the vast studied topics over the past decades in quantum mechanics is to get an approximate or exact solution of the Schrodinger, Dirac, and Klein-Gordon equation with some typical potential \cite{gc.1,gc.2,gc.3,gc.4,gc.5} using various methods. These methods are Nikiforov-Uvarov (NU) method \cite{gc.6,gc.6a}, Nikiforov–Uvarov-Functional-Analysis (NUFA) method \cite{nufa,nufa.1},  parametric Nikiforov-Uvarov (pNU) method \cite{ibo,ibo.1},   asymptotic iteration method (AIM) \cite{gc.7}, super symmetric quantum mechanics (SUSYQM) \cite{gc.8,gc.9}, supersymmetric WKB (SWKB) approach \cite{eomu}  factorization method  \cite{gc.10,gc.11}, the exact quantization rule method \cite{gc.12} and proper quantization rule \cite{gc.13}, the Laplace transformation method (LTM)\cite{gc.13a} artificial neural network method (ANN) \cite{gc.13aa} and the analytical exact iterative method (AEIM) \cite{gc.13aaa} etc.

In recent times, Due to the numerous discoveries of excited states of either heavy or heavy-light mesons \cite{gc.13aaaa} and the noteworthy experimental progress in understanding meson spectroscopy have piloted the exaggerated study of meson spectroscopy and proliferated in the theoretical model \cite{gc.13aaaaa}. for the role, the quark model has acclaimed considerable success in the quark model \cite{gc.13a1}. Quark and its anti-quark are the constituents of the mesons which is a harmonic sub particle \cite{gc.13a2}. Various hypothetical models \cite{gc.13a3} have assimilated the spectroscopy assignment of a heavy-light meson (HLM). The theory of quantum chromodynamics (QCD) is to understand the dynamics and behavior of heavy mesons \cite{gc.13a3,gc.13a5}. For a similar measure heavy-light flavor mesons spectroscopy \cite{gc.13a3} can be directed by QCD\cite{gc.13a6}. Kwong et al. \cite{gc.13a7,gc.13a8,gc.13a9,gc.13a10}, Godfrey and Rosner \cite{gc.13a11,gc.13a12,gc.13a13}, eichten et al. \cite{gc.13a14} and Barnes and Godfray \cite{gc.13a15} highlighted this heavy-light transition. Much more radially and orbitally excited states, however, require more experimental research\cite{gc.13a16}.  The Regge trajectories, spectroscopy, and decay properties of $B$ and $B_{s}$ mesons \cite{gc.13aaaaa} $D$ and $D_{s}$ mesons \cite{v.kher}, as well as radiative transitions and mixing parameters of the meson have been explored using the Gaussian wave function with quark-antiquark potential model \cite{gc.13a18}. Hassanabadi et al. \cite{gc.13a19} used the variational approach to solve the radial SE and looked at the parameters of light and heavy mesons while considering Cornell interaction. Using a relativistic quark model based on a quasi-potential approach, Ebert et al. \cite{gc.13a20} investigated the mass spectra and radiative decay rates of charmonium, bottomonium, and $b\bar{c}$ mesons. Wong \cite{gc.13a21} used a temperature-dependent Yukawa potential obtained from lattice gauge computations to compute the dissociation temperature of heavy quarkonia. For deducing the distinct characteristics of HLM, the mass spectra of the quarkonium system as charm and bottom mesons with quark and anti-quark interaction potential \cite{gc.13a22,gc.13a23,gc.13a24} by using the NU method. Al-Jamel and Widyan \cite{gc.13a25} have correlated the spin-average mass spectra of heavy quarkonia with coulomb plus quadratic potential. The mass and leptonic decay width of $c\bar{c},b\bar{b},c\bar{s},b\bar{s},b\bar{u},c\bar{b}$ has been numerically computed by about salen \cite{gc.13a27} using the Jacobi method. Using the field correlation method decay constant of HLM has been calculated \cite{gc.13a28}.  Our doubly flavored hydraulic system is a combination set of heavy-light (KS/KQ; K=C, B) mesons. Properties of HLM were apprehended by various attempts using heavy quark as non-relativistically and light relativistically \cite{gc.13a29,gc.13a30}.
E. Omugbe et al. \cite{eom.1} examined Any $\ell$
-State Energy of the Spinless Salpeter Equation Under the Cornell Potential by the WKB Approximation Method: An Application to Mass Spectra of Mesons
Abdelmadjid \cite{abd} studied Relativistic Interactions in One-Electron Atoms with Modified Yukawa Potential for Spin $\frac{1}{2}$ Particles. Maireche A. A new study of relativistic and nonrelativistic for new modified Yukawa potential via the BSM in the framework of noncommutative quantum mechanics symmetries: An application to heavy-light mesons systems. \cite{abd.1}. A Quarkonium potential is solved within the nonrelativistic regime using the Nikiforov-Uvarov method to obtain the energy spectrum and the wave function in terms of Laguerre polynomials studied by E.P. Inyang \cite{epi}.  In the realm of relativistic and non-relativistic phenomena, numerous research seems to entail their interest with a combination of two or more potentials with an intention of a broader passel of applications \cite{gc.13a31,gc.13a32}. Relativistic and non-relativistic thermal properties with bound and scattering states of the Klein-Gordon equation for Mobius square plus generalized Yukawa potentials studied by A. N. Ikot et al. \cite{ani}.  E. Omugbe et al. \cite{eom} recently studied Approximate mass spectra and root mean square radii of quarkonia using Cornell potential plus spin-spin interactions. Inyang, E.P. et al.\cite{iep} investigate the Masses and thermal properties of a Charmonium and Bottomonium Mesons. Mass spectrum of heavy quarkonium for screened Kratzer potential using the series expansion method studied by E. E. Ibekwe et al.\cite{eei}.  E.Omugbe et al. \cite{eo} recently investigate Approximate mass spectra and root mean square radii of quarkonia using Cornell potential plus spin-spin interactions. Study of $B$, $ B_{s}$ mesons using heavy quark effective theory studied by K. Gandhi \cite{kg.1}. In the field of meson spectroscopy, despite the significant correction term in the potentials or the uses of a different combination of potentials, there was a low mass spectra theory \cite{v.patel}. In the HLM, the observed mass of newly excited states was less than the experimental data or their perspective of other theoretical study data \cite{gc.13a33,gc.13a34,gc.13a35}. This was then taken into account by a confining term to be played in the role along with coloumbic terms in the combinations of different potentials. A compelling suppression of mass spectra for $n\geq3$  was observed and defined by the selection of factor as screening part in potential \cite{v.patel1}. Franz Gross et.al presents a comprehensive review of both the theory and experimental successes of Quantum Chromodynamics and includes a review of the earliest theoretical and experimental foundations \cite{f.g}. Hua-Xing Chen et.al provides an updated review of the recent experimental and theoretical progresses \cite{h.x}

Through various experimental Collaborations, many newly exciting states of heavy-light flavor charm mesons are found such as Belle \cite{gc.14a1,gc.14a2}, LHCb \cite{gc.14a3,gc.14a4,gc.14a5,gc.14a6}, BESIII \cite{gc.14a7}, BABAR \cite{gc.14a8,gc.14a9,gc.14a10,gc.14a11} and upcoming experimental facility like will also have charm physics program \={P}ANDA \cite{pa.1,pa.2,pa.3,pa.4,pa.5,pa.6,pa.7}.

Pretentious from the above studies, we further attempt to investigate the behavior of the quarkonium system by selecting two different phenomenological potentials \cite{gc.13a36} (liner plus modified Yukawa potential) and fitting the spectra with coefficient potentials.

Linear plus modified Yukawa potential (LIMYP) takes as
\begin{equation} V(r)=A_{1} r+ \frac{A_{2} e^{-\alpha r}}{r}-\frac{A_{3} e^{-2 \alpha r}}{r^2}+A_{4} \label{eq:gc1x}. \end{equation}
where

 $A_{1},A_{2},A_{3}$ and $A_{4}$ are potential strength and $\alpha$ is screening parameter

\begin{equation}e^{-\alpha r}=1-\frac{\alpha r}{1!}+\frac{\alpha^2 r^2}{2!}-\frac{\alpha^3 r^3}{3!} \label{eq:gc1xa}. \end{equation}
\begin{equation} e^{-2 \alpha r }= 1-\frac{2 \alpha r}{1!}+\frac{4 \alpha^2 r^2}{2!}-\frac{8 \alpha^3 r^3}{3!} \label{eq:gc1xb}. \end{equation}

We carry out a series expansion of the exponential term in Eq. (\ref{eq:gc1xa}) and Eq. (\ref{eq:gc1xb}) up to order three, in the model the potential to interact in the quark-antiquark system and substitute the results into Eq.(\ref{eq:gc1x}) this yields.

\begin{equation} V(r)=\frac{\eta_{1}}{r^2} + \frac{\eta_{2}}{r}+\eta_{3}r +\eta_{4}r^2+\eta_{5} \label{eq:gc3}.  \end{equation}

where
\[ \eta_{1}= -A_{3}  ,\eta_{2}=A_{2} + 2 A_{3} \alpha, \eta_{3}= A_{1}+\frac{A_{2}\alpha^2}{2}+\frac{4}{3}A_{3}\alpha^3 \]
\begin{equation} \eta_{4}= -\frac{A_{2} \alpha^3 }{6}, \eta_{5}=-A_{2} \alpha - 2 A_{3} \alpha^2+A_{4}\label{eq:gc3a}.\end{equation}

The third term of Eq. (\ref{eq:gc3}) is a linear term for confinement feature and the second term is coulomb potential which describes the short distance between quarks.

We organized our work into two parts. In the first part of this paper, sec.2 and sec.3 review the theory of the generalized NU method, and in the latter, we obtain the energy eigenvalue and corresponding normalized wave function of LIMYP. In the second part, from sec.4 and sec.5, we obtain the mass spectra of LIMYP using the non-relativistic quark model and in the latter second part, we present the results and discussion of mass spectra of all combination sets of HLM, along with the spectra of parameters involved in energy eigenvalue. While we summarize our conclusion in the final section.

\section{ Nikiforov-Uvarov (NU) method }

The NU method was proposed by Nikiforove and Uvarov \cite{gc.6,gc.13a37} to transform a Schrodinger-like equation into a second-order differential equation via a coordinate transformation $ q=q(r)$,of the form
\begin{equation}  \psi^{''}(q)+\frac{\tilde{\tau} (x)}{\sigma (q)} \psi^{'}(q)+\frac{\tilde{\sigma}(q)}{\sigma^{2}(q)}\psi(q)=0 \label{eq:gc1}. \end{equation}
 where $\tilde{\sigma}(q)$ and $\sigma(q)$ are polynomials, at the most second degree, and $\tilde{q}$ is a first-degree polynomial.The exact solution of Eqs.(\ref{eq:gc1}) can be obtained by using the transformation.
  \begin{equation}  \psi(q)=\phi(q)y(q) \label{eq:gc1h}.  \end{equation}
 This transformation reduce Eqs.(\ref{eq:gc1}) into a hypergeometric type Eqs.(6) into hypergeometric-type equation of the form

 \begin{equation}  \sigma(q)y^{''}(q)+\tau(q)y{'}(q)+\lambda y(q)=0. \end{equation}

  The function $\phi(q)$ can be defined as the logarithm derivative
 \begin{equation} \frac{\phi^{'}(q)}{\phi (q)}=\frac{\pi(q)}{\sigma (q)} \label{eq:gc1d}. \end{equation}

  With $\pi(q)$ being at most a first-degree polynomial. The second part $\psi(q)$ begin y(q) in Eqs.(7) is the hypergeometric function with the polynomial solution given by Rodrigues relation as
  \begin{equation}  y(q)=\frac{B_{nl}}{\rho(q)}\frac{d^{n}}{dq^{n}}\left[\sigma^{n}(q)\rho(x)\right] \label{eq:gc1e}.\end{equation}
   where
   $B_{nl}$ is the normalization constant and $\rho(x) $ the weight function which satisfied the condition below;
   \begin{equation} \left(\sigma(q)\rho(q)\rho(q)\right)^{'}=\tau(q)\rho(q) \label{eq:gc1f}.\end{equation}
  where also
  \begin{equation}  \tau(x)=\tilde{\tau}(x)+2\pi(x)\label{eq:gc1b}. \end{equation}

  For bound state solutions, it required that
 \begin{equation}  \tau^{'}(q)<0.\end{equation}
   The eigenfunction  and eigenvalues can be obtained using the definition of the following function $\pi(q)$ and parameter $\lambda$, respectively:
  \begin{equation}  \pi(q)= \frac{\sigma^{'}(q)-\tilde{\tau}(q)}{2}\pm\sqrt{\left(\frac{\sigma^{'}(q)-\tilde{\tau}(q)}{2}\right)^2- \tilde{\sigma}(q)+k\sigma(q)} \label{eq:gc2}.  \end{equation}

   and

   \begin{equation}  \lambda=k_{-}+\pi_{-}^{'}(q) \label{eq:gc2c}. \end{equation}

   The value of $k$ can be obtained by setting the discriminant in the square root Eq. (\ref{eq:gc2}) equal to zero. As such, the new eigenvalues equation can be given as

   \begin{equation}  \lambda+n\tau^{'}(q)+\frac{n(n-1)}{2}\sigma^{''}(q)=0, (n=0,1,2,....) \label{eq:gc2d}. \end{equation}

\section{ Bound state solution of the Klein-Gordon equation with linear plus modified Yukawa potential (LIMYP)}
 The Klein-Gordon equation for a spinless particle for $\hbar=c=1$ in N-dimensions is given as \cite{gc.35}
   \[ \left[-\nabla^{2}+\left(M+S(r)\right)^2+\frac{(N+2\ell-1)(N+2\ell-3)}{4 r^2}\right]\psi(r,\theta,\varphi)=\]\begin{equation}[E_{n\ell}-V(r)]^2 \psi(r,\theta,\varphi).\end{equation}

  where $\nabla^2$ is Laplacian, $M$ is the reduced mass, $E_{n\ell}$ is the energy spectrum $n$ and $\ell$ are the radial and orbital angular momentum quantum number respectively. It is well known that for the wave function to satisfy the boundary condition it can be rewritten as

  \begin{equation}\psi(r,\theta,\varphi)=\frac{R_{n\ell}}{r}Y_{\ell m}(\theta, \varphi).\end{equation}
  The angular component of the wave function could be separated leaving only the radial part as shown below

\begin{eqnarray}
\label{eq:gc4}
\nonumber 
  & \frac{d^2 R(r)}{dr^2} & +[(E_{n \ell}^2-M^2)+V^2(r)-S^2(r)-2(E_{n \ell}V(r)+M S(r)) \\
  &-&\frac{(N+2 \ell-1)(N+2 \ell-3)}{4r^2}] R(r)=0.
\end{eqnarray}
 Thus, for equal vector and scalar potentials $V(r)=S(r)=2V(r)$, then eq. (\ref{eq:gc4}) becomes

    \begin{eqnarray}
     \nonumber 
     &\frac{d^2 R(r)}{dr^2}&+\left[(E_{n \ell}^2-M^2)-2V(r)(E_{n \ell} + M)-\frac{(N+2 \ell-1)(N+2 \ell-3)}{4r^2}\right]\\
      &R(r)&=0 \label{eq:gc5}.
      \end{eqnarray}

    Upon substituting Eq.(\ref{eq:gc3}) into Eq.(\ref{eq:gc5}), we obtain

\begin{eqnarray}
 \nonumber
& \frac{d^2 R(r)}{dr^2}+  \left[(E_{n \ell}-M^2)+(-\frac{ 2\eta_{1}}{r^2} + \frac{2 \eta_{2}}{r}- 2 \eta_{3}r+ 2\eta_{4}r^2- 2\eta_{5})(E_{n \ell} +M)\right] R(r) &\\
&-\left[\frac{(N+2 \ell-1)(N+2 \ell-3)}{4 r^2}\right]R(r)=0 & \label{eq:gc6}.  \end{eqnarray}

In order to transform the coordinate from $r$ to $q$ in Eq. (\ref{eq:gc6}), we set

\begin{equation} q=\frac{1}{r} \label{eq:gc7}. \end{equation}

  This implies that $2^{nd}$ derivative in Eqs (\ref{eq:gc7}) becomes;
  \begin{equation}  \frac{d^2 R(r)}{dr^2}= 2q^3 \frac{dR(q)}{dq}+q^4\frac{d^2 R(q)}{dq^2} \label{eq:gc8}. \end{equation}

  Substituting Eqs.(\ref{eq:gc7}) and (\ref{eq:gc8}) into Eq. (\ref{eq:gc6})

 \[\frac{d^2 R(q)}{dq^2}+\frac{2}{q}\frac{dR}{dq}+\frac{1}{q^4}\]
 \[\left[(E_{n \ell}^2-M^2)+\left(-2 \eta_{4} r^2+2 \eta_{3}r- \frac{2 \eta_{2}}{r}+\frac{2 \eta_{1}}{r^2}-2 \eta_{5}\right)(E_{n \ell}+M)\right]R(q)\]
 \begin{equation}\left[-\frac{(N+2 \ell-1)(N+2 \ell-3)x^2}{4}\right]R(q)=0 \label{eq:gc8c}. \end{equation}

   Next, we propose the following approximation scheme on the term $\frac{\eta_{1}^{'}}{q^{2}} $ and  $\frac{\eta_{2}^{'}}{q} $.

   Let us assume that there are characteristics radius $r_{0}$ of meson. Then scheme is based on the expansion of $\frac{\eta_{1}^{'}}{q^{2}} $ and  $\frac{\eta_{2}^{'}}{q} $ in power series arond $r_{0}$; i.e., around $\delta\equiv\frac{1}{r_{0}}$, in the X-space up to second order.This is similar to the Pekeris approximation, Which helps to deform the centrifugal term such that the potential can be solved by the NU method \cite{gc.35}.

   Setting $y=q-\delta$ and around $y=0$, it can be expanded into a series of powers as;

  \begin{equation}\frac{\eta_{2}}{q}=\frac{\eta_{2}}{y+\delta}=\frac{\eta_{2}}{\delta\left(1+\frac{y}{\delta}\right)}=\frac{\eta_{2}}{\delta}\left(1+\frac{y}{\delta}\right)^{-1}.\end{equation}

    which yields

 \begin{equation}  \frac{\eta_{2}}{q}= \eta_{2} \left(\frac{3}{\delta}-\frac{3 x}{\delta^2}+\frac{x^2}{\delta^2}\right)\label{eq:gc8a}. \end{equation}

  similarly,

\begin{equation}\frac{\eta_{1}}{q^{2}}= \eta_{1}\left(\frac{6}{\delta^2}-\frac{8x}{\delta^3}+\frac{3 x^2}{\delta^4}\right) \label{eq:gc8b}. \end{equation}

  By substituting Eqs.(\ref{eq:gc8a}) and (\ref{eq:gc8b}) into Eq.(\ref{eq:gc8c}), we obtain

 \begin{equation} \frac{d^2 R(q)}{dq^2}+\frac{2q}{q^2}\frac{dR(q)}{dq}+\frac{1}{4}\left[-\varepsilon+\eta q-\gamma q^2 \right]R(q)=0 \label{eq:gc9}. \end{equation}

where

\begin{equation} -\varepsilon = (E_{n \ell}^2-M^2)-\frac{6 \eta_{3}}{\delta}(E_{n \ell}+M)+\frac{12 \eta_{4}}{\delta^2}(E_{n \ell}+M)-2 \eta_{5}(E_{n \ell}+M). \end{equation}
\begin{equation} \eta= 2 \eta_{3}(E_{n \ell}+M)+\frac{6 \eta_{3}}{\delta^2}(E_{n \ell}+M)-\frac{16 \eta_{4}}{\delta^3}(E_{n \ell}+M).\end{equation}
\begin{equation}\gamma= 2 \eta_{1}(E_{n \ell}+M)+\frac{2 \eta_{3}}{\delta^3}(E_{n \ell}+M)-\frac{6 \eta_{4}}{\delta^4}(E_{n \ell}+M)+\frac{(N+2 \ell-1)(N+2 \ell-3)}{4}. \end{equation}

Comparing Eq.(\ref{eq:gc9}) and Eq.(\ref{eq:gc1}) we obtain

\begin{equation} \tilde{\tau}(q)=2q ,\ \ \ \   \sigma(q)=q^2 \label{eq:gc10}.\end{equation}
\begin{equation} \tilde{\sigma}(q)=-\varepsilon+\eta q-\gamma q^2 \label{eq:gc11}.\end{equation}
\begin{equation} \sigma^{'}(q)=2 q, \ \ \ \  \sigma^{''}(q)=2 \label{eq:gc12}.\end{equation}
We substitute Eq. (\ref{eq:gc10}), (\ref{eq:gc11}),(\ref{eq:gc12}) and Eq.(\ref{eq:gc2}) and obtain
\begin{equation} \pi(q)=\pm \sqrt{\varepsilon-\eta q +(\gamma+k) q^2} \label{eq:gc13}. \end{equation}

 To determine $k$, we take the discriminant of the function under the square root, which yields

\begin{equation} k=\frac{\eta^2-4 \gamma \varepsilon}{4 \varepsilon} \label{eq:gc14}. \end{equation}

 We substitute Eq.(\ref{eq:gc14}) into Eq.(\ref{eq:gc13}) and have

\begin{equation} \pi(q)=\pm \left(\frac{\eta q}{2 \sqrt{\varepsilon}}-\frac{\varepsilon}{\sqrt{\varepsilon}}\right)\label{eq:gc15}.  \end{equation}

 We take the negative part of Eq.(\ref{eq:gc15}) and differentiate, which yields
\begin{equation} \pi_{-}^{'}(q)=-\frac{\eta}{2 \sqrt{\varepsilon}} \label{eq:gc16}.  \end{equation}

 By substituting Eqs.(\ref{eq:gc10}), (\ref{eq:gc11}),(\ref{eq:gc12}) and (\ref{eq:gc16}  ) into Eq.(\ref{eq:gc1b}) we have

 \begin{equation} \tau(q)= 2q-\frac{\eta x}{\sqrt{\varepsilon}}+\frac{2 \varepsilon}{\sqrt{\varepsilon}} \label{eq:gc17}.  \end{equation}
 Differentiating Eq.(\ref{eq:gc17}) we have

 \begin{equation} \tau^{'}(q)=2-\frac{\eta}{ \sqrt{\varepsilon}}. \end{equation}

 By using Eq.( \ref{eq:gc2c} ) we obtain

\begin{equation}  \lambda=\frac{\eta^2-4 \gamma \varepsilon}{4 \varepsilon}-\frac{\eta}{2 \sqrt{\varepsilon}}\label{eq:gc18}. \end{equation}
 And using Eq.( \ref{eq:gc2d} ), we obtain

\begin{equation} \lambda_{n}=\frac{n \eta}{\sqrt{\varepsilon}}-n^2-n \label{eq:gc19}. \end{equation}

 Equating Eqs.(\ref{eq:gc18}) and (\ref{eq:gc19}), and substituting Eqs.(\ref{eq:gc3a}) and (28), yields the energy eigenvalue equation of the LIMYP in the relativistic limit as

 Considering a transformation of the form: $M+E_{n \ell}\rightarrow \frac{2 \mu}{\hbar^2}$ and $ M-E_{n \ell} \rightarrow -E_{n \ell} $ where $\mu$ is reduced mass. We have the non-relativistic energy eigenvalues equation as,

\[ E_{n \ell}=-\frac{(6 A_{1}+A_{2} \alpha^2+4 A_{3} \alpha^3)}{\delta}-\frac{A_{2} \alpha^3}{\delta^2}-2(A_{4}-2 A_{3}\alpha^3- A_{2} \alpha)-\frac{\hbar^2}{8 \mu}\]
\begin{equation}\left[\frac{\frac{4\mu}{\hbar^2}(A_{2}+2 A_{3} \alpha)+\frac{2 \mu}{\hbar^2 \delta^2}(6A_{1}+A_{2}\alpha^2+4 A_{3}\alpha^3)-\frac{16 \mu A_{2} \alpha^3}{3 \hbar^2 \delta^3}}{n+\frac{1}{2}+\sqrt{\frac{1}{4}+\frac{4 A_{3}\mu}{\hbar^2}+\frac{4\mu}{\hbar^2 \delta^3}(A_{1}+\frac{A_{2}\alpha^2}{2}+\frac{4}{3}A_{3} \alpha^3)+\frac{A_2 \alpha^3 \mu}{\hbar^2 \delta^4}+\frac{(N+2 \ell-1)(N+2 \ell-3)}{4}}}\right]^2 \label{eq:gc.19}.\end{equation}

 To determine the wavefunction, we substitute Eqs., (\ref{eq:gc10}),(\ref{eq:gc11}) (\ref{eq:gc12}) and (\ref{eq:gc15}) into Eq.(\ref{eq:gc1d}) and obtain

 \begin{equation} \frac{d\phi}{\phi}=\left(\frac{\varepsilon}{q^2\sqrt{\varepsilon}}-\frac{\alpha}{2q\sqrt{\varepsilon}}\right)dq \label{eq:gc20}. \end{equation}

Integrating Eq.(\ref{eq:gc20}),we obtain
\begin{equation} \phi(q)= q^{-\frac{\alpha}{2\sqrt{\varepsilon}}} e^{-\frac{\varepsilon}{q\sqrt{\varepsilon}}} \label{eq:gc21}. \end{equation}

By substituting Eqs.(\ref{eq:gc10}),(\ref{eq:gc11}) (\ref{eq:gc12}) and (\ref{eq:gc1d}) into Eq.(\ref{eq:gc1f})and integrating,thereafter simplify we obtain

\begin{equation} \rho(q)= q^{-\frac{\alpha}{\sqrt{\varepsilon}}} e^{-\frac{2 \varepsilon}{q\sqrt{\varepsilon}}}. \end{equation}

Substituting Eqs. (21) and (42) into Eq.(10) we have

\begin{equation} y_{n}(q)=B_{n}e^{\frac{2\varepsilon}{q\sqrt{\varepsilon}}}q^{\frac{\alpha}{\sqrt{\varepsilon}}}\frac{d^{n}}{dq^{n}}\left[e^{-\frac{2\varepsilon}{x\sqrt{\varepsilon}}}q^{2n-\frac{\alpha}{\sqrt{\varepsilon}}}\right]\end{equation}

Rodrigue's formula of the associated Laguerre polynomials is

\begin{equation} L_{n}^{\frac{\alpha}{\sqrt{\varepsilon}}}\left[\frac{2\varepsilon}{q\sqrt{\varepsilon}}\right]=\frac{1}{n!}e^{\frac{2\varepsilon}{q\sqrt{\varepsilon}}}q^{{\frac{\alpha}{\sqrt{\varepsilon}}}}\frac{d^n}{dq^n}\left(e^{-\frac{2\varepsilon}{q\sqrt{\varepsilon}}}q^{2n-\frac{\alpha}{\sqrt{\varepsilon}}}\right).\end{equation}

where
\begin{equation} \frac{1}{n!}=B_{n}  \label{eq:gc22}. \end{equation}

 Hence,

\begin{equation}y_{n}(q)\equiv L_{n}^{\frac{\alpha}{\sqrt{\varepsilon}}}\left(\frac{2\varepsilon}{q\sqrt{\varepsilon}}\right).\end{equation}

Substituting Eqs.(\ref{eq:gc21}) and (\ref{eq:gc22}) into Eq.(\ref{eq:gc1h}) we obtain the wavefunction of Eq.(\ref{eq:gc6}) in terms of Lagurre polynomial as

\begin{equation} \psi(q)=B_{nl}q^{-\frac{\alpha}{2\sqrt{\varepsilon}}}e^{-\frac{\varepsilon}{q\sqrt{\varepsilon}}}L_{n}^{\frac{\alpha}{\sqrt{\varepsilon}}}\left(\frac{2\varepsilon}{q\sqrt{\varepsilon}}\right).\end{equation}

where$ N_{nl}$ is the normalization constant, Which can be obtained from

\begin{equation}\int_{0}^{\infty}|B_{nl}(r)|^2 dr=1. \end{equation}

\section{Mass spectra}
We drive the mass spectra of the heavy-light mesons system such as $c\bar{s}$, $c\bar{q}$, $b\bar{s}$,  and $b\bar{q}$  that have the quark and antiquark flavor. To determine the mass spectra we use the following relation.

\begin{equation} M=m_{1}+m_{2}+E_{n \ell} \label{eq:gc.22a}.\end{equation}



substituting Eq.(\ref{eq:gc.22a}) into Eq.(\ref{eq:gc.19}) we obtain
\[M=m_{1}+m_{2}+ \left(-\frac{(6 A_{1}+A_{2} \alpha^2+4 A_{3} \alpha^3)}{\delta}\right)-\frac{A_{2} \alpha^3}{\delta^2}-2(A_{4}-2 A_{3}\alpha^3- A_{2} \alpha)-\frac{\hbar^2}{8 \mu}\]
\begin{equation}\left[\frac{\frac{4\mu}{\hbar^2}(A_{2}+2 A_{3} \alpha)+\frac{2 \mu}{\hbar^2 \delta^2}(6A_{1}+A_{2}\alpha^2+4 A_{3}\alpha^3)-\frac{16 \mu A_{2} \alpha^3}{3 \hbar^2 \delta^3}}{n+\frac{1}{2}+\sqrt{\frac{1}{4}+\frac{4 A_{3}\mu}{\hbar^2}+\frac{4\mu}{\hbar^2 \delta^3}(A_{1}+\frac{A_{2}\alpha^2}{2}+\frac{4}{3}A_{3} \alpha^3)+\frac{A_2 \alpha^3 \mu}{\hbar^2 \delta^4}+\frac{(N+2 \ell-1)(N+2 \ell-3)}{4}}}\right]^2 \label{eq:gc23}.\end{equation}

\vspace{5pt}
\begin{table}[h!]
\begin{center}
\caption{The mass spectra of  $c\bar{s}$ for parameters $m_c=1.209 GeV$,$m_{\bar{s}}=0.586 GeV$, $\hbar=1$ $N=3$,$\mu=0.394693$, $\delta=5.8532$,  $A_1=3.289 GeV$, $A_2=1.90892217 GeV $, $A_3=1.90896574 GeV$, $A_4=0.000241 GeV$  }
\begin{tabular}{ |c|c|c|c|c|c|c|c|  }
\hline
State   &   Our work  &   \cite{v.patel}   & \cite{s.godfrey}  &  \cite{v.kher} &  \cite{n.devlani} & \cite{d.ebert}  & Exp\cite{zyla} \\ [1ex]

\hline
1S &  2.076 & 2.076 & 2.091 & 2.076 &2.076& 2.075&2.076 \\ [1ex]
\hline
1P   &  2.515 & 2.534 & 2.563 &2.542 & 2.540 & 2.537&2.514\\ [1ex]
\hline
2P    &   3.019 & 3.015 & 3.034 & 3.090 &  3.026 &  3.119&-\\ [1ex]
\hline
2S  &  2.636 &  2.673 & 2.717 & 2.709 & 2.713&  2.720&2.609\\ [1ex]
\hline
3S  &  3.061  & 3.130 &  3.222 & 3.261 & 3.175 & 3.236&- \\ [1ex]

\hline
4S  & 3.244 &  3.500 & 3.568 & 3.772 & 3.567 &  3.664&-\\ [1ex]

\hline
1D  &  2.831 & 2.844  & 2.912 & 2.871 & 2.852 &  2.950&-\\ [1ex]

\hline
2D &  3.112 &  3.262 & 3.310 & 3.383 & 3.277 &  3.436&- \\ [1ex]
\hline
1F  &  3.083 &  3.095 &- &-&-&-&-\\ [1ex]

\hline
\end{tabular}
\end{center}
\end{table}

\vspace{5pt}
\begin{table}[h!]
\begin{center}
\caption{The mass spectra of  $c\bar{q}$ for parameters for parameters $m_c=1.209 GeV$,$m_{\bar{q}}=0.46 Gev$, $\hbar=1$ $N=3$, $\mu=0.333217$, $\delta=5.8102$,  $A_1=3.289 GeV$, $A_2=1.90892217 GeV $, $A_3=1.90896574 GeV$, $A_4=0.000241 GeV$ }
\begin{tabular}{ |c|c|c|c|c|c|c|c|  }
\hline


State   &   Our work  & \cite{v.patel1}   & \cite{s.godfrey}  &  \cite{v.kher} &  \cite{n.devlani1} & \cite{d.ebert} &Exp \cite{c.patrigani}  \\ [1ex]
\hline
1S &  1.978 &1.975 & 2.000 &  1.975 &  1.973 & 1.975&1.975\\ [1ex]

\hline
1P   &  2.434 &   2.448 & 2.473 & 2.440 &2.448 & 2.414&2.434\\ [1ex]
\hline
2P    &   2.953 &  2.977 &  2.948 &  3.027 &  2.949 & 2.986&-\\ [1ex]
\hline
2S  &  2.665 & 2.624 & 2.628 & 2.636 &  2.586 &  2.619&2.613\\ [1ex]
\hline
3S  &  3.074   & 3.118 & 3.100&  3.225 & 3.104 & 3.087&-\\ [1ex]

\hline
4S  & 3.341 &  3.512 & 3.490 & 3.778& 3.510 & 3.474&-\\ [1ex]

\hline
1D  &  2.783 & 2.777 & 2.830 &  2.779 &  2.768 & 2.834&-\\ [1ex]

\hline
2D &  3.132&  3.242 & 3.229 & 3.338 & 3.207 & 3.293&-\\ [1ex]
\hline
1F  &  3.009 &  3.048 & - &- &-&-&-\\ [1ex]
\hline
\end{tabular}
\end{center}
\end{table}

\vspace{5pt}
\begin{table}[h!]
\begin{center}
\caption{The mass spectra of  $b\bar{s}$ for parameters $m_b=4.823 GeV$,$m_{\bar{s}}=0.586 Gev$, $\hbar=1$ $N=3$
,   $\mu=0.52251$, $\delta=2.5288$, $A_1=1.000 GeV$, $A_2=1.83972217 GeV $, $A_3=1.30956574 GeV$, $A_4=0.000241 GeV$}
\begin{tabular}{ |c|c|c|c|c|c|c|c|  }
\hline
State   &   Our work  &   \cite{v.patel2}  &  \cite{gc.13aaaaa}  &  \cite{d.ebert} & \cite{j.liu} &\cite{m.shah} & Exp\cite{PDG}       \\  [1ex]

\hline
1S &  5.401 & 5.401 &  5.401 &  5.404 & 5
370&5.403&5.401 \\  [1ex]

\hline
1P   & 5.850  &   5.838 & 5.835 & 5.844 & 5.838&5.838&- \\  [1ex]
\hline
2P    &   6.380 &    6.316  &  6.380 & 6.343 & 6.254&6.233&- \\  [1ex]
\hline
2S  &6.168 & 5.990 & 6.023& 5.988  & 5.971&5.952&- \\  [1ex]
\hline
3S  &  6.544   & 8.443 &  6.570&  6.473 & - &6.425&-  \\  [1ex]

\hline
4S  &  6.756 & 6.788 &  7.083 &  6.878 &  -&6.863 &- \\  [1ex]

\hline
1D  & 6.179 & 6.139 &  6.150  &   6.200 & 6.117&6.181&-\\  [1ex]

\hline
2D &  6.604 &  6.560 & 6.668 &  6.635   &  6.450&6.626&- \\  [1ex]

\hline
1F  &  6.570 & 6.385 & - &   -  & - &-&-\\   [1ex]
\hline
\end{tabular}
\end{center}
\end{table}

\vspace{5pt}
\begin{table}[h!]
\begin{center}
\caption{The mass spectra of  $b\bar{q}$ for parameters $m_b=4.823 GeV$, $m_{\bar{q}}=0.46 Gev$, $\hbar=1$ $N=3$, $\mu=0.4199469$, $\delta=2.5288$, $A_1=1.000GeV$, $A_2=1.50992217 GeV $, $A_3=1.31196574 GeV$, $A_4=0.002241GeV$}
\begin{tabular}{ |c|c|c|c|c|c|c|c|  }
\hline


State   &   Our work  &   \cite{v.patel2}  &  \cite{gc.13aaaaa}  &  \cite{d.ebert} & \cite{j.liu}&\cite{m.shah} & Exp\cite{PDG} \\ [1ex]
\hline
1S &  5.314 & 5.314 &  5.314 &  5.314 & 5.288&5.314 & 5.314 \\  [1ex]
\hline
1P   &  5.747  &   5.779 & 5.740 & 5.745 & 5.759&5.737&- \\ [1ex]
\hline
2P    &   6.100 &    6.307  &  6.301 & 6.249 & 6.188&6.127&- \\ [1ex]
\hline
2S  & 5.924 & 5.951 & 5.942 & 5.902  & 5.903&5.819&- \\ [1ex]
\hline
3S  &  6.214  & 6.425 &  6.504 &  6.385 &- &6.251 &-  \\ [1ex]
\hline

4S  &6.474 & 6.846 &  6.772 &6.785 &  - &6.647 &- \\ [1ex]
\hline
1D  &6.035 & 6.104 &  6.057  &   6.106 & 6.042&6.065&-\\ [1ex]
\hline
2D &6.273 &  6.571 & 6.596 &  6.540   & 6.377&6.429 &- \\ [1ex]
 \hline
1F  &6.252 & 6.343 & - &   -  & - &-&-\\ [1ex]
 \hline
\end{tabular}
\end{center}
\end{table}

\begin{figure}[!tbp]
  \centering
  \begin{minipage}[b]{0.45\textwidth}
\centerline { \includegraphics[width=1.2\textwidth]{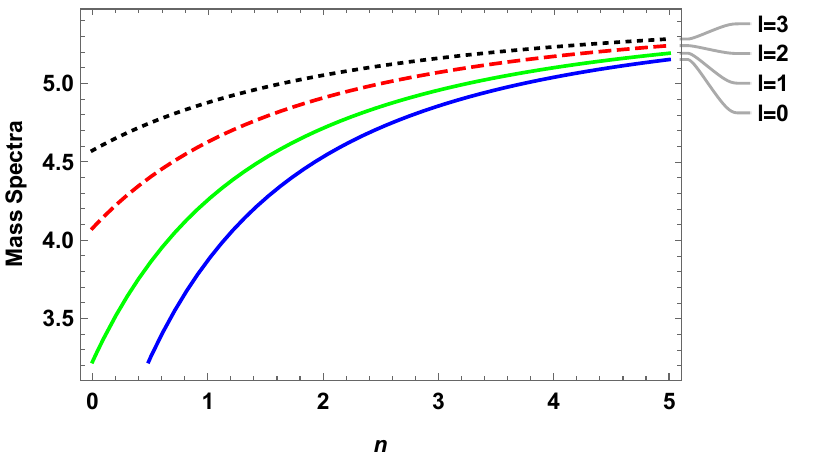}}
  \caption{ Variation of mass spectra of $c\bar{s}$ with principal quantum number($n$) .}\label{fig:cs.1}
 \end{minipage}
  \hfill
  \begin{minipage}[b]{0.45\textwidth}
\centerline { \includegraphics[width=1.2\textwidth]{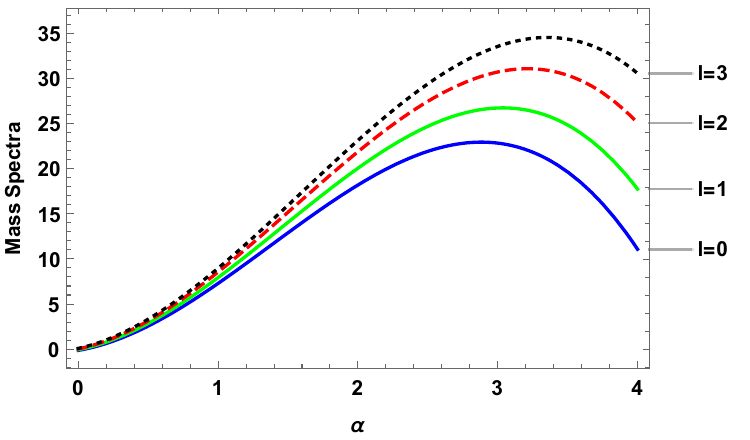}}
  \caption{Variation of mass spectra of $c\bar{s}$ with screening parameter $\alpha$ with different azimuthal quantum number ($\ell$).}\label{fig:cs.2}
 \end{minipage}
\end{figure}

\begin{figure}[!tbp]
  \centering
  \begin{minipage}[b]{0.45\textwidth}
\centerline { \includegraphics[width=1.2\textwidth]{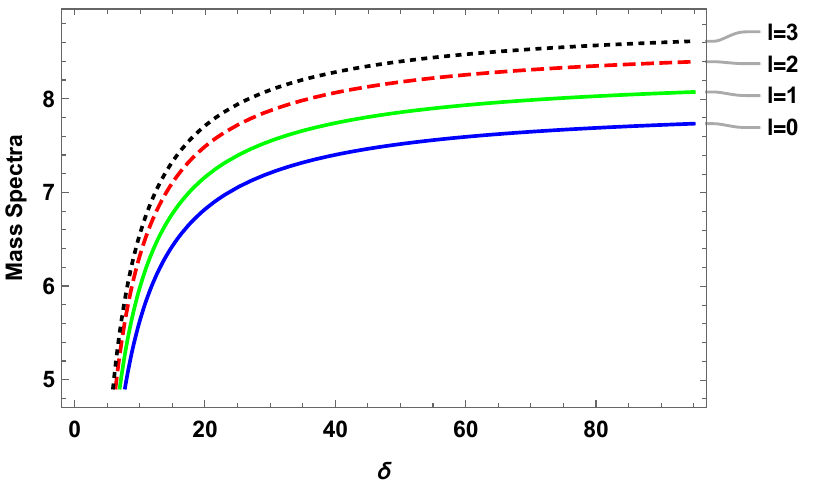}}
  \caption{ Variation of mass spectra of $c\bar{s}$ with $\delta$ for different azimuthal quantum number ($\ell$).}\label{fig:cs.3}
 \end{minipage}
  \hfill
  \begin{minipage}[b]{0.45\textwidth}
 \centerline {\includegraphics[width=1.2\textwidth]{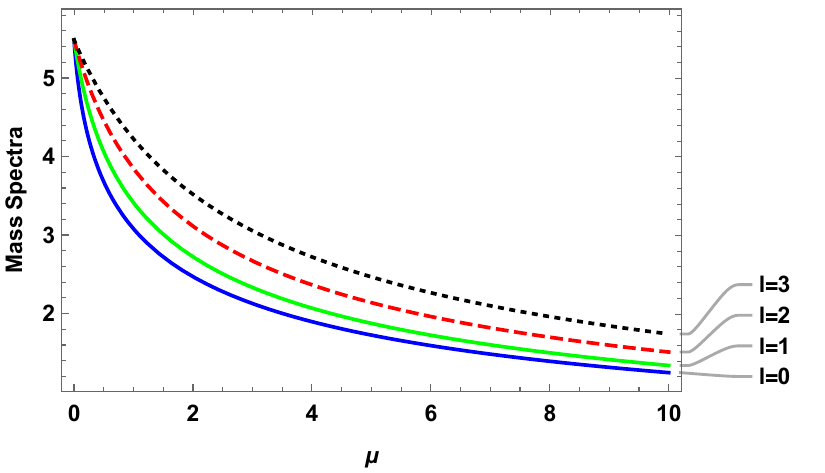}}
  \caption{Variation of mass spectra of $c\bar{s}$ with reduced mass ($\mu$) for different azimuthal quantum number ($\ell$).}\label{fig:cs.4}
 \end{minipage}
\end{figure}

\begin{figure}[!tbp]
  \centering
  \begin{minipage}[b]{0.45\textwidth}
 \centerline {\includegraphics[width=1.2\textwidth]{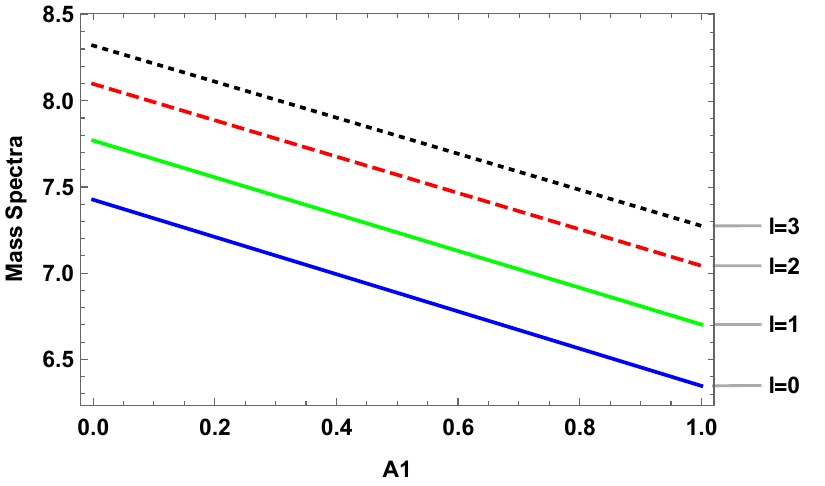}}
  \caption{ Variation of mass spectra of $c\bar{s}$ with potential strength($A_1$) for different azimuthal quantum number ($\ell$).}\label{fig:cs.5}
 \end{minipage}
  \hfill
  \begin{minipage}[b]{0.45\textwidth}
\centerline { \includegraphics[width=1.2\textwidth]{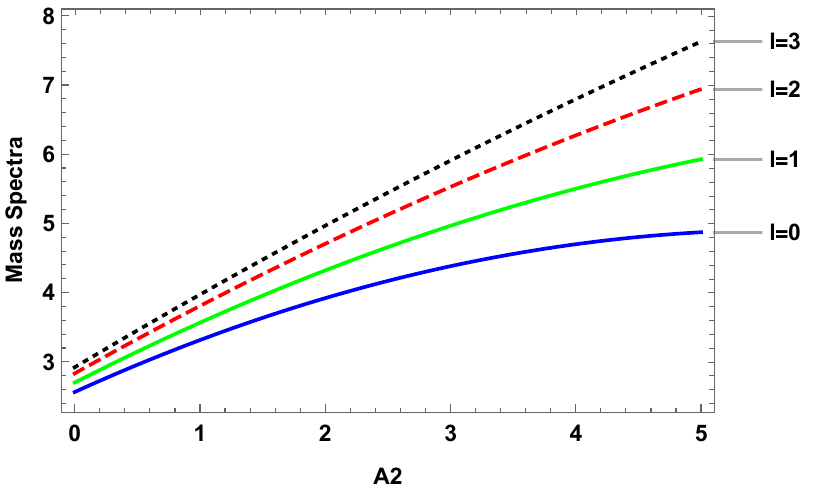}}
  \caption{Variation of mass spectra of $c\bar{s}$ with  potential strength($A_2$) for different azimuthal quantum number ($\ell$).}\label{fig:cs.6}
 \end{minipage}
\end{figure}

\begin{figure}[!tbp]
  \centering
  \begin{minipage}[b]{0.45\textwidth}
\centerline { \includegraphics[width=1.2\textwidth]{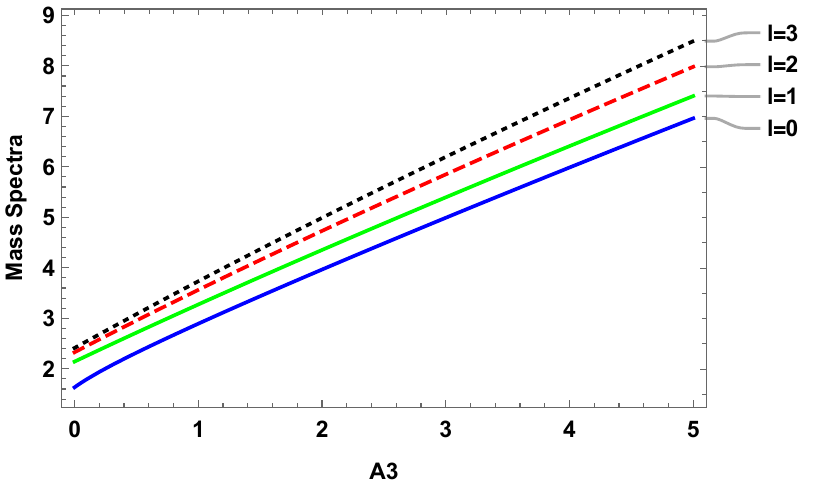}}
  \caption{ Variation of mass spectra of $c\bar{s}$ with potential strength($A_3$) for different azimuthal quantum number ($\ell$).}\label{fig:cs.7}
 \end{minipage}
  \hfill
  \begin{minipage}[b]{0.45\textwidth}
\centerline { \includegraphics[width=1.2\textwidth]{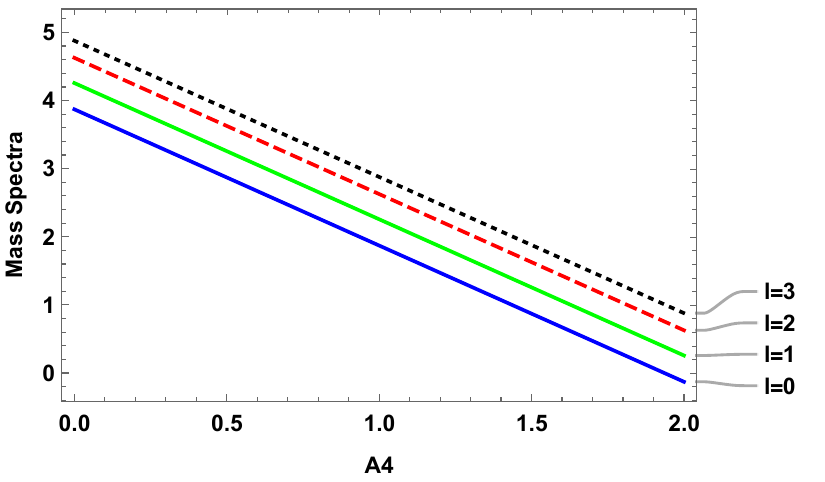}}
  \caption{Variation of mass spectra of $c\bar{s}$ with potential strength($A_4$) for different azimuthal quantum number ($\ell$).}\label{fig:cs.8}
 \end{minipage}
\end{figure}

\begin{figure}[!tbp]
  \centering
  \begin{minipage}[b]{0.45\textwidth}
\centerline { \includegraphics[width=1.2\textwidth]{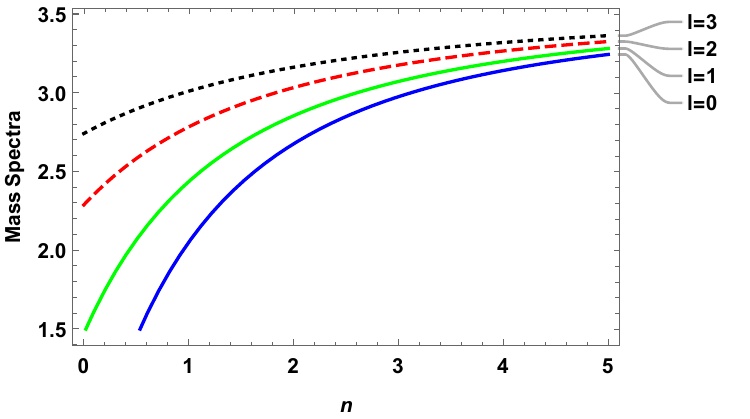}}
  \caption{ Variation of mass spectra of $c\bar{q}$ with principal quantum number($n$) .}\label{fig:cq.1}
 \end{minipage}
  \hfill
  \begin{minipage}[b]{0.45\textwidth}
\centerline { \includegraphics[width=1.2\textwidth]{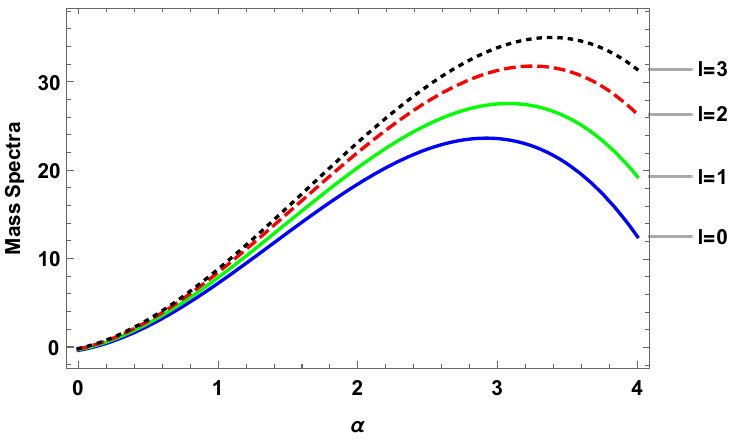}}
  \caption{Variation of mass spectra of $c\bar{q}$ with screening parameter $\alpha$ with different azimuthal quantum number ($\ell$).}\label{fig:cq.2}
 \end{minipage}
\end{figure}

\begin{figure}[!tbp]
  \centering
  \begin{minipage}[b]{0.45\textwidth}
\centerline { \includegraphics[width=1.2\textwidth]{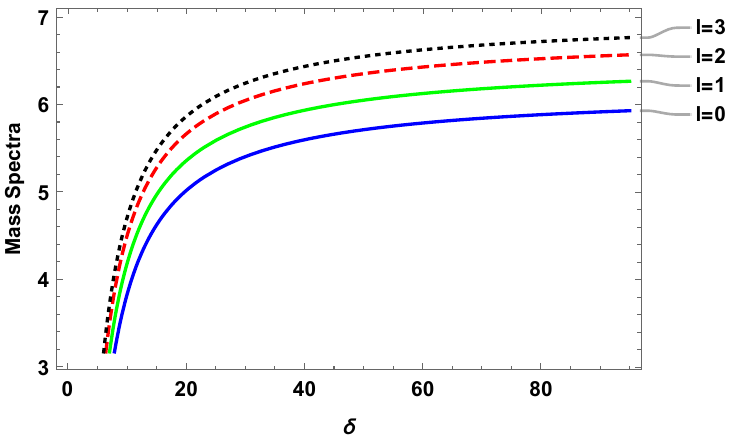}}
  \caption{ Variation of mass spectra of $c\bar{q}$ with $\delta$ for different azimuthal quantum number ($\ell$).}\label{fig:cq.3}
 \end{minipage}
  \hfill
  \begin{minipage}[b]{0.45\textwidth}
\centerline { \includegraphics[width=1.2\textwidth]{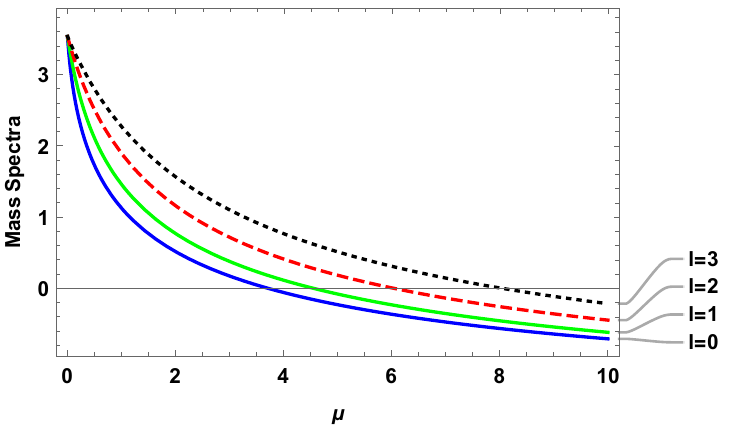}}
  \caption{Variation of mass spectra of $c\bar{q}$ with reduced mass ($\mu$) for different azimuthal quantum number ($\ell$).}\label{fig:cq.4}
 \end{minipage}
\end{figure}

\begin{figure}[!tbp]
  \centering
  \begin{minipage}[b]{0.45\textwidth}
\centerline { \includegraphics[width=1.2\textwidth]{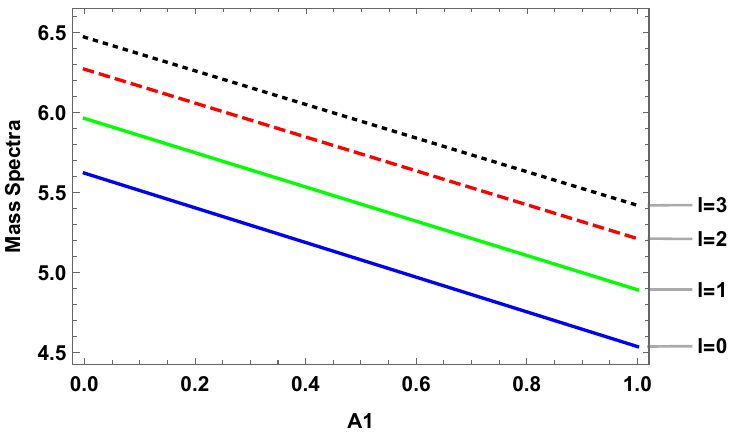}}
  \caption{ Variation of mass spectra of $c\bar{q}$ with potential strength($A_1$) for different azimuthal quantum number ($\ell$).}\label{fig:cq.5}
 \end{minipage}
  \hfill
  \begin{minipage}[b]{0.45\textwidth}
\centerline { \includegraphics[width=1.2\textwidth]{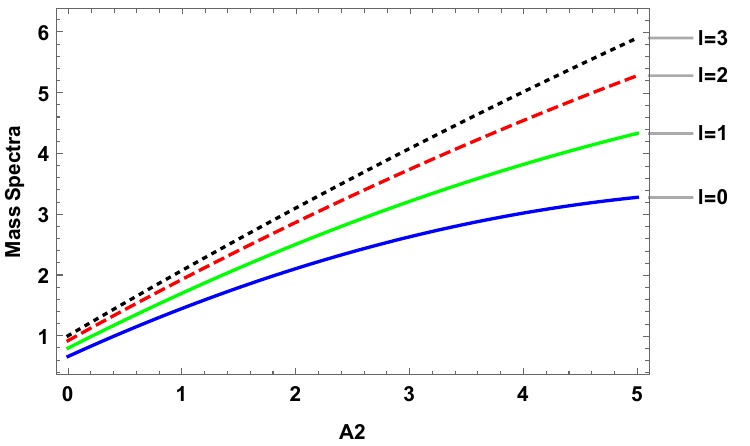}}
  \caption{Variation of mass spectra of $c\bar{q}$ with potential strength($A_2$) for different azimuthal quantum number ($\ell$).}\label{fig:cq.6}
 \end{minipage}
\end{figure}

\begin{figure}[!tbp]
  \centering
  \begin{minipage}[b]{0.45\textwidth}
 \centerline{\includegraphics[width=1.2\textwidth]{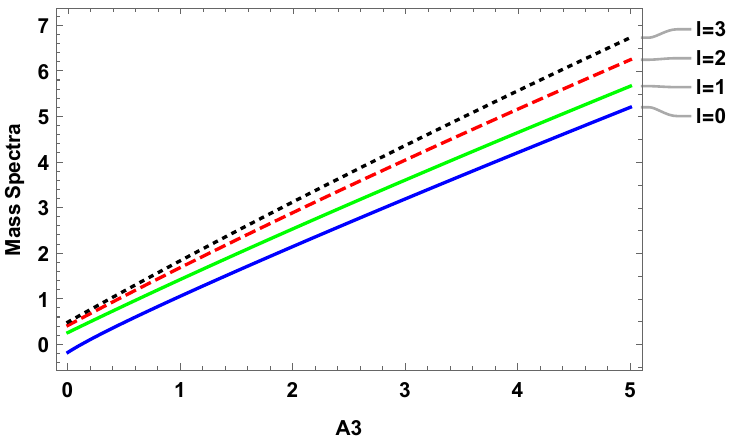}}
  \caption{ Variation of mass spectra of $c\bar{q}$ with potential strength($A_3$) for different azimuthal quantum number ($\ell$).}\label{fig:cq.7}
 \end{minipage}
  \hfill
  \begin{minipage}[b]{0.45\textwidth}
\centerline { \includegraphics[width=1.2\textwidth]{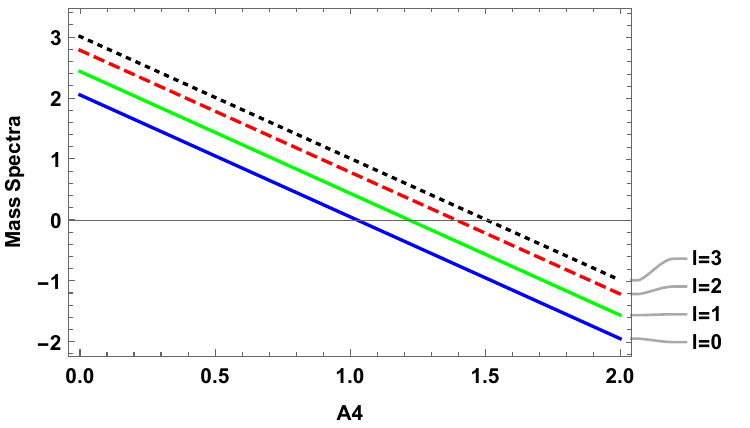}}
  \caption{Variation of mass spectra of $c\bar{q}$ with potential strength($A_4$) for different azimuthal quantum number ($\ell$).}\label{fig:cq.8}
 \end{minipage}
\end{figure}

\begin{figure}[!tbp]
  \centering
  \begin{minipage}[b]{0.45\textwidth}
\centerline { \includegraphics[width=1.2\textwidth]{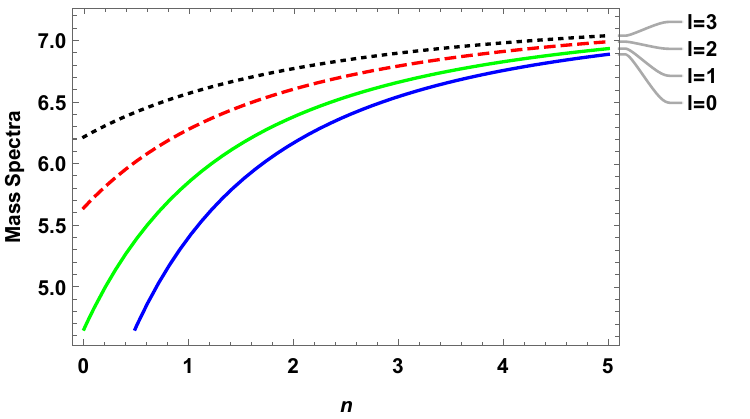}}
  \caption{ Variation of mass spectra of $b\bar{s}$ with principal quantum number($n$) .}\label{fig:bs.1}
 \end{minipage}
  \hfill
  \begin{minipage}[b]{0.45\textwidth}
\centerline { \includegraphics[width=1.2\textwidth]{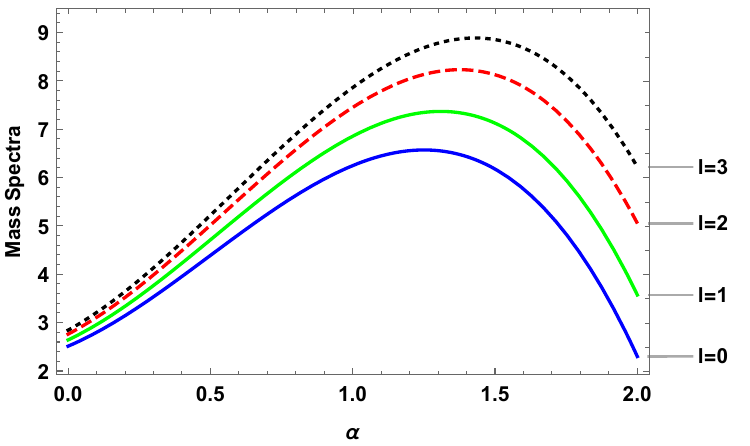}}
  \caption{Variation of mass spectra of $b\bar{s}$ with screening parameter $\alpha$ with different azimuthal quantum number ($\ell$).}\label{fig:bs.2}
 \end{minipage}
\end{figure}

\begin{figure}[!tbp]
  \centering
  \begin{minipage}[b]{0.45\textwidth}
\centerline { \includegraphics[width=1.2\textwidth]{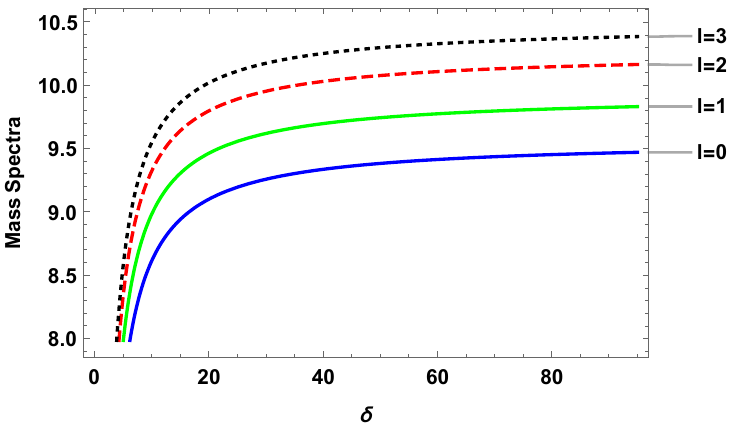}}
  \caption{ Variation of mass spectra of $b\bar{s}$ with $\delta$ for different azimuthal quantum number ($\ell$).}\label{fig:bs.3}
 \end{minipage}
  \hfill
  \begin{minipage}[b]{0.45\textwidth}
\centerline { \includegraphics[width=1.2\textwidth]{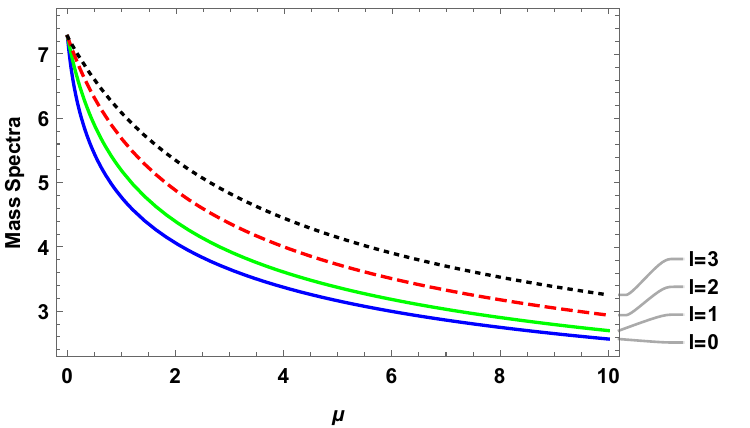}}
  \caption{Variation of mass spectra of $b\bar{s}$ with reduced mass ($\mu$) for different azimuthal quantum number ($\ell$).}\label{fig:bs.4}
 \end{minipage}
\end{figure}

\begin{figure}[!tbp]
  \centering
  \begin{minipage}[b]{0.45\textwidth}
\centerline { \includegraphics[width=1.2\textwidth]{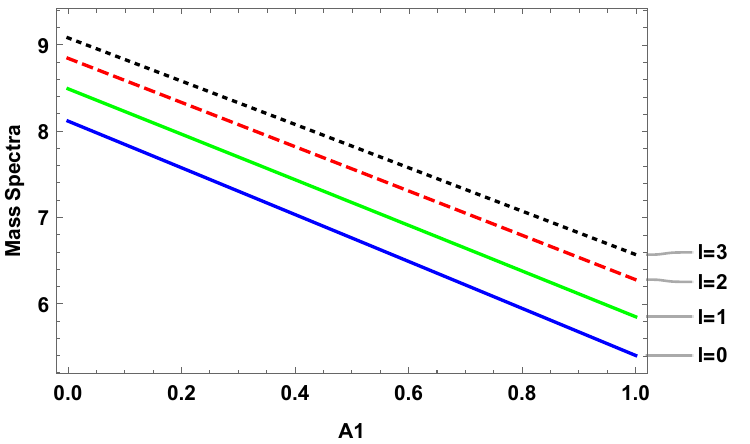}}
  \caption{ Variation of mass spectra of $b\bar{s}$ with potential strength($A_1$) for different azimuthal quantum number ($\ell$).}\label{fig:bs.5}
 \end{minipage}
  \hfill
  \begin{minipage}[b]{0.45\textwidth}
\centerline { \includegraphics[width=1.2\textwidth]{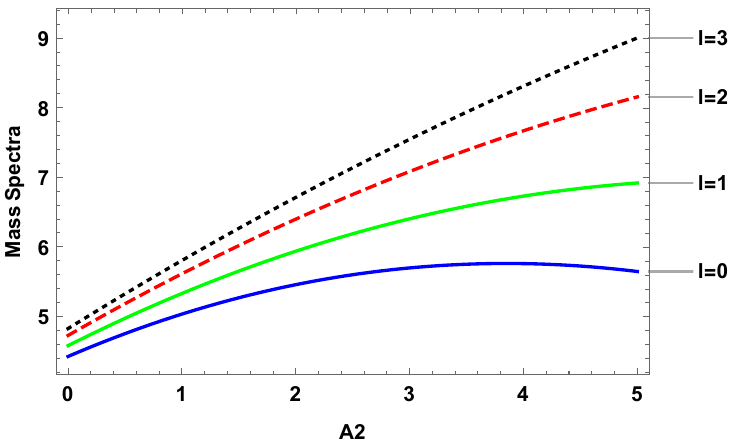}}
  \caption{Variation of mass spectra of $b\bar{s}$ with reduced with potential strength($A_2$) for different azimuthal quantum number ($\ell$).}\label{fig:bs.6}
 \end{minipage}
\end{figure}

\begin{figure}[!tbp]
  \centering
  \begin{minipage}[b]{0.45\textwidth}
\centerline { \includegraphics[width=1.2\textwidth]{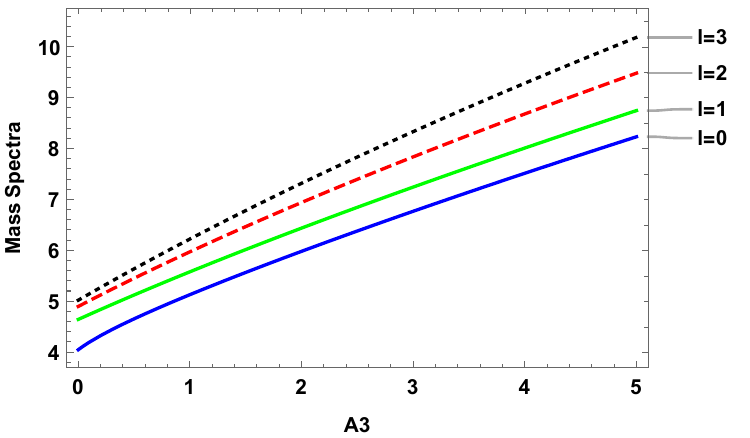}}
  \caption{ Variation of mass spectra of $b\bar{s}$ with potential strength($A_3$) for different azimuthal quantum number ($\ell$).}\label{fig:bs.7}
 \end{minipage}
  \hfill
  \begin{minipage}[b]{0.45\textwidth}
\centerline { \includegraphics[width=1.2\textwidth]{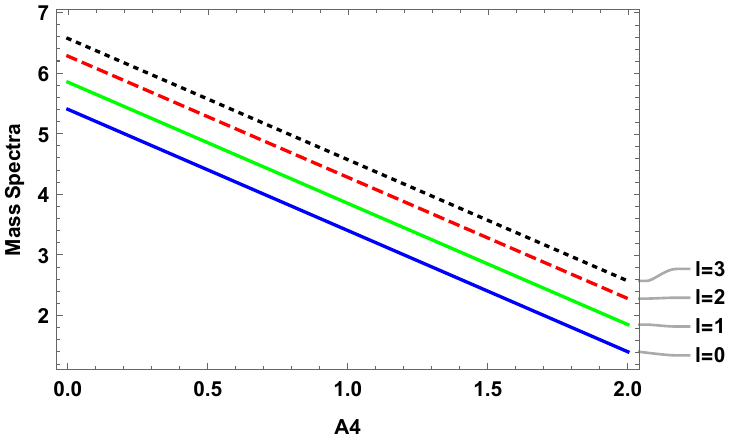}}
  \caption{Variation of mass spectra of $b\bar{s}$ with reduced with potential strength($A_4$) for different azimuthal quantum number ($\ell$).}\label{fig:bs.8}
 \end{minipage}
\end{figure}

\begin{figure}[!tbp]
  \centering
  \begin{minipage}[b]{0.45\textwidth}
\centerline { \includegraphics[width=1.2\textwidth]{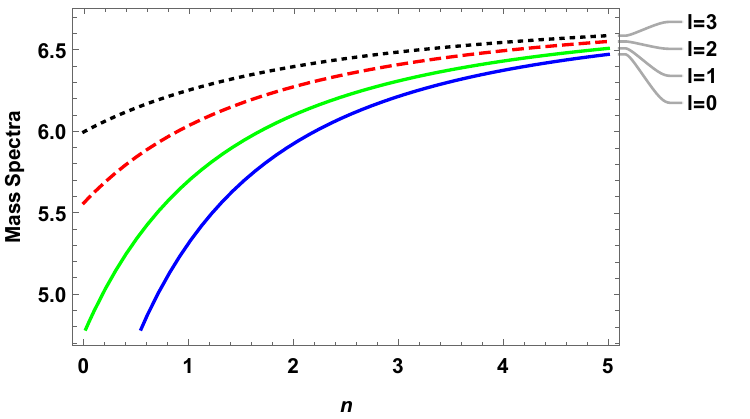}}
  \caption{ Variation of mass spectra of $b\bar{q}$ with principal quantum number($n$) .}\label{fig:bq.1}
 \end{minipage}
  \hfill
  \begin{minipage}[b]{0.45\textwidth}
\centerline { \includegraphics[width=1.2\textwidth]{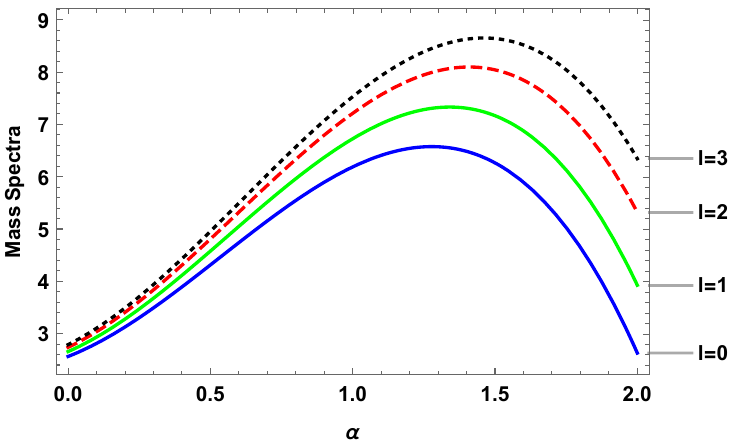}}
  \caption{Variation of mass spectra of $b\bar{q}$ with screening parameter $\alpha$ with different azimuthal quantum number ($\ell$).}\label{fig:bq.2}
 \end{minipage}
\end{figure}

\begin{figure}[!tbp]
  \centering
  \begin{minipage}[b]{0.45\textwidth}
\centerline { \includegraphics[width=1.2\textwidth]{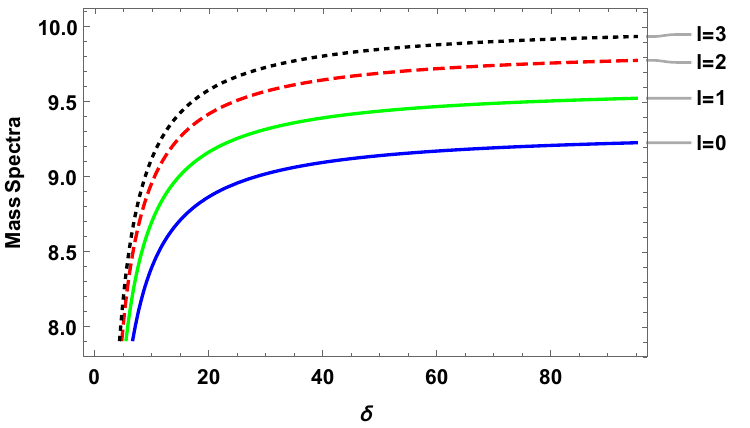}}
  \caption{ Variation of mass spectra of $b\bar{q}$ with $\delta$ for different azimuthal quantum number ($\ell$).}\label{fig:bq.3}
 \end{minipage}
  \hfill
  \begin{minipage}[b]{0.45\textwidth}
\centerline { \includegraphics[width=1.2\textwidth]{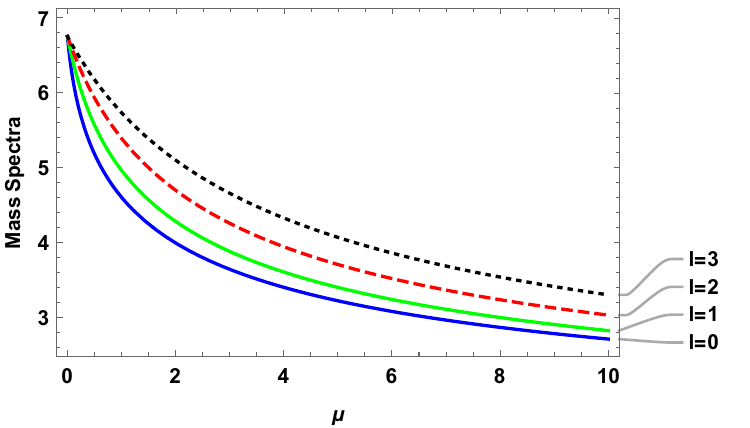}}
  \caption{Variation of mass spectra of $b\bar{q}$ with reduced mass ($\mu$) for different azimuthal quantum number ($\ell$).}\label{fig:bq.4}
 \end{minipage}
\end{figure}

\begin{figure}[!tbp]
  \centering
  \begin{minipage}[b]{0.45\textwidth}
\centerline { \includegraphics[width=1.2\textwidth]{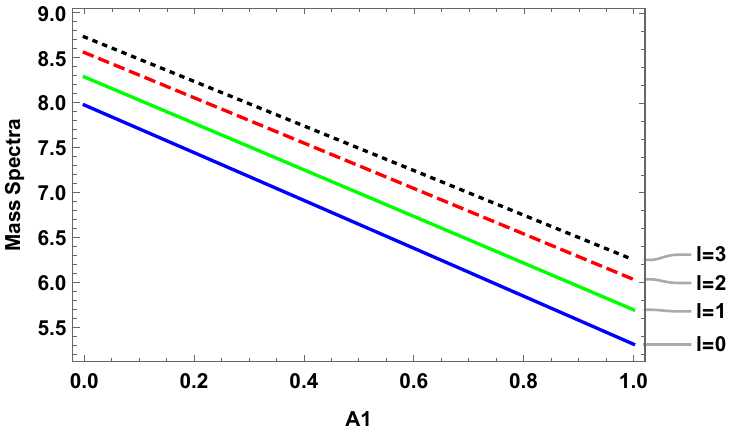}}
  \caption{ Variation of mass spectra of $b\bar{q}$ with potential strength($A_1$) for different azimuthal quantum number ($\ell$).}\label{fig:bq.5}
 \end{minipage}
  \hfill
  \begin{minipage}[b]{0.45\textwidth}
\centerline { \includegraphics[width=1.2\textwidth]{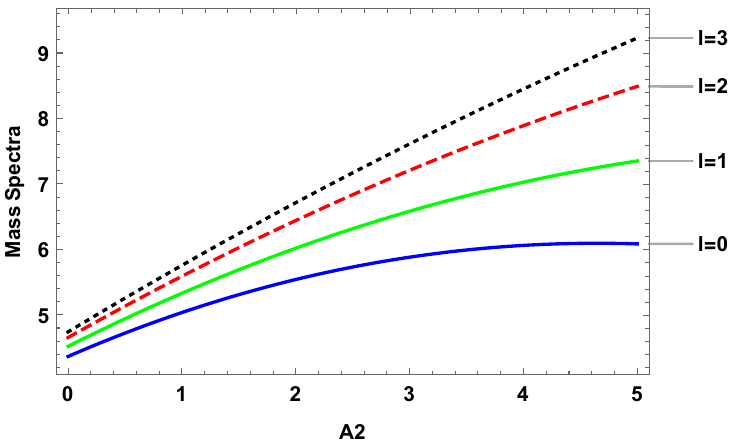}}
  \caption{Variation of mass spectra of $b\bar{q}$  with potential strength($A_2$) for different azimuthal quantum number ($\ell$).}\label{fig:bq.6}
 \end{minipage}
\end{figure}

\begin{figure}[!tbp]
  \centering
  \begin{minipage}[b]{0.45\textwidth}
 \centerline{\includegraphics[width=1.2\textwidth]{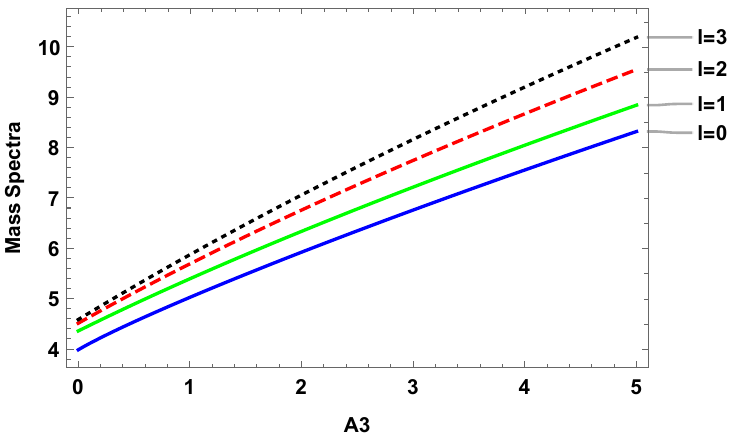}}
  \caption{ Variation of mass spectra of $b\bar{q}$ with potential strength($A_3$) for different azimuthal quantum number ($\ell$).}\label{fig:bq.7}
 \end{minipage}
  \hfill
  \begin{minipage}[b]{0.45\textwidth}
\centerline { \includegraphics[width=1.2\textwidth]{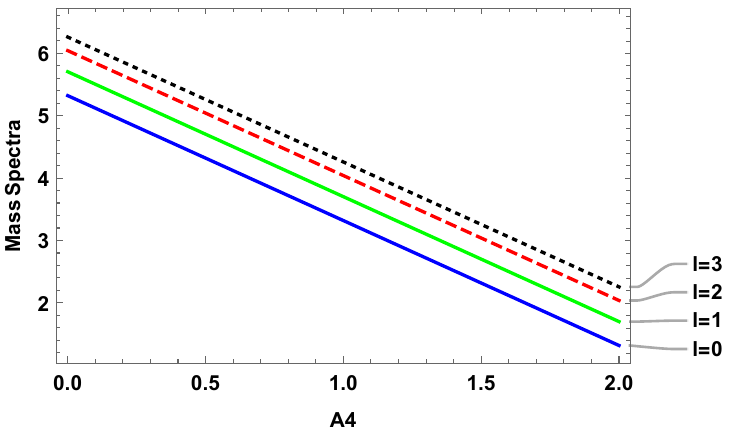}}
  \caption{Variation of mass spectra of $b\bar{q}$ with potential strength($A_4$) for different azimuthal quantum number ($\ell$).}\label{fig:bq.8}
 \end{minipage}
\end{figure}

\section{Results and discussion}
The mass spectra obtained by combination sets of a heavy-light meson (HLM) flavor mesons from the bound state solution of the K-G equation which is a combination of linear plus modified Yukawa potential for states ranging from 1S, 2S, 3S, 4S, 1P, 2P, 1D, 2D, and 1F. The four HLMs combination mass spectra $c\bar{s}$, $c\bar{q}$, $b\bar{s}$, $b\bar{q}$ are listed in Tables 1, 2, 3, and 4 respectively. Mass spectra of all four HLMs combination sets are relatively accurate when compared to both experimental data and other theoretical calculations. The current potential framework’s results are highly consistent with the experimental data of various states to date. 1S, 2S, and 1P are the only states for the $c\bar{s}$ combination that has been experimentally verified that have mass spectra difference by 0.0 MeV, 1.0 MeV, and 2.7 MeV with a \% error of 0\%, 0.03\%, and 1.02\%. For $c\bar{q}$, combination 1s,1p, and 2s are the only states known experimentally and found to have mass spectra differences of 0MeV, 3.0 MeV, 0.0 MeV, and 5.2 MeV with errors of 0.15\%, 0\%, and 1.95\%. While for $b\bar{q}$, and $b\bar{s}$ combinations the only experimental state known is 1s, and in both cases, we found a 0 MeV difference with 0 \% error. All other states of the corresponding combination sets of HLM are in good agreement with other theoretical investigations, along with the experimental results of a few states. I.e., in higher quantum states mass spectra have been somehow less than in other theoretical studies. For $n\geq3$ we found effective screening as a factor coming in our combined potential term. From the possible effective screening factor in the energy eigenvalues eq (54), it is possible to understand the low mass spectra theory of higher states. The energy eigenvalues equation derived from the combined potential (1) comprises several variables that, when plotted against mass spectra, provide a more detailed understanding of the potential under consideration. It is possible to comprehend the plot of mass spectra with part 1 variables like the principal number (n), reduced mass ($\mu$), alpha ($\alpha$), and delta ($\delta$) for all corresponding combination sets of HLMs ($c\bar{s}$, $c\bar{q}$, $b\bar{s}$, $b\bar{q}$). For the principal quantum number(n) in Fig: 1, 9, 17, 25, the plot mimics the root function plot to represent the calculated data and converges at higher quantum numbers, which can be attributed to an increase in screening power with increasing quantum numbers. While in Fig: 2 10, 18, 26, alpha ($\alpha$) mimics a cubic function plot that grows initially before decreasing. While in fig:3, 11, 19, and 27 mimic a log function plot and show an asymptotic convergence towards zero for the delta ($\delta$), which increases with an increase in the mass spectrum. In Fig: 4, 12, 20, 28, reduced mass ($\mu$), mass spectra decrease as reduced mass is increased, mimicking a reciprocal plot and producing an asymptotic convergence away from the origin. From part 2 variables, A1 and A4 variables exhibit a negative linear plot, fig: 5, 8, 13, 16, 21, 24, 29, 32. In other words, the mass spectra decline as A1 and A4 levels rise. A2 and A3 show an opposite trend in the plot from A1 and A4, i.e., mass spectra increase with the increase in A2 and A3 and further diverge in the plot with the increase in the potential value of A2 and A3, respectively fig: 6, 7, 14, 15, 22, 23, 30, 31, for all respective permuted combinations of heavy-light mesons.

\section{ Conclusion}
We have successfully calculated the mass spectra of all heavy-light mesons (HLMs) combination sets utilizing the combined potential framework, where the potential was used to derive the energy eigenvalue using the linear plus modified Yukawa potential. The mass spectra data has a high degree of similarity when compared to experimental data that is currently known, as well as the minimum \% error when compared to other theoretical study data. We can see this effect in HLMs where the higher state mass spectra are suppressed due to the screening effect with few MeV with experimental as well as other theoretical results. A good picture of the potential employed in this work may be seen in the plots of mass spectra concerning variables. This study has demonstrated the significance of non-relativistic correction in a proposed model for accurate spectroscopic parameter prediction for the $c\bar{s}$, $c\bar{q}$, $b\bar{s}$, $b\bar{q}$ mesons. To quantify this HLM precisely in the future, more experimental work will be required. A forthcoming experimental facility \={P}ANDA and another experimental facility like BABAR, Belle, and LHCb will be in a special position for that. The potential also utilized in this study can also be useful in nuclear particle physics (decay properties for heavy-heavy and heavy-light mesons), atomic and molecular physics, hot and dense QCD media, etc.


\end{document}